%% file: EML_ReRevised_Unmarked.tex
\documentclass[3p]{elsarticle}
\usepackage{epstopdf}
\usepackage{subfigure,subcaption,graphicx,float}
\usepackage{amsmath, amsfonts, amssymb}
\usepackage{amsthm}
\usepackage{yhmath}
\usepackage{epsf}
\usepackage{amsfonts}
\usepackage{stmaryrd}
\usepackage{amssymb}
\usepackage{leftidx}
\usepackage{xcolor}
\usepackage{mathtools}
\usepackage{placeins}
\usepackage{booktabs}
\usepackage{enumitem}
\usepackage{caption}
\usepackage{tabu}
\usepackage{multirow}
\usepackage{stackengine}
\usepackage[utf8]{inputenc}
\usepackage[english]{babel}
\usepackage{bm}
\usepackage{array}
\usepackage{multirow}

\input{definitions}

\theoremstyle{plain}

\long\def\symbolfootnote[#1]#2{\begingroup%
\def\thefootnote{\fnsymbol{footnote}}\footnote[#1]{#2}\endgroup}

\begin{document}
\begin{frontmatter}

\title{Why planar cracks fragment into echelon cracks \vspace{0.1cm}}

\vspace{-0.1cm}

\author[mech]{Olivia Ward}
\ead{oward31@gatech.edu}

\author[civil]{Aditya Kumar\corref{cor1}}
\ead{aditya.kumar@ce.gatech.edu}

\address[mech]{George W. Woodruff School of Mechanical Engineering, Georgia Institute of Technology, Atlanta, GA 30332, USA 
\vspace{0.05cm}}

\address[civil]{School of Civil and Environmental Engineering, Georgia Institute of Technology, Atlanta, GA 30332, USA \vspace{0.05cm}}

\cortext[cor1]{Corresponding author}

\begin{abstract}

\vspace{-0.1cm}

Predicting the path and shape of growing cracks is fundamental to understanding of fracture.
Under out-of-plane shear loading, an initially planar crack may spontaneously fragment into multiple cracks, forming a striking echelon crack pattern. Explaining this crack morphogenesis in brittle materials has been a long-standing open problem essential to developing a complete theory of crack growth.
Here, through comparison with classical experiments, we show that a strength-constrained minimization of the sum of elastic and surface energies explains echelon crack formation. Results are presented for both soft and hard materials, confirming the model's general applicability to any brittle material. As a corollary, we show that, contrary to prevailing views, a purely energetic minimization model is insufficient to predict the growth of large cracks. We identify two key non-dimensional parameters governing crack fragmentation and orientation, and demonstrate that these reconcile the various energy-based and stress-based empirical criteria proposed in the literature for crack path.


\end{abstract}

\end{frontmatter}

\section{Introduction}

\begin{figure*}[t]
	\centering
	\includegraphics[width=1.0\textwidth]{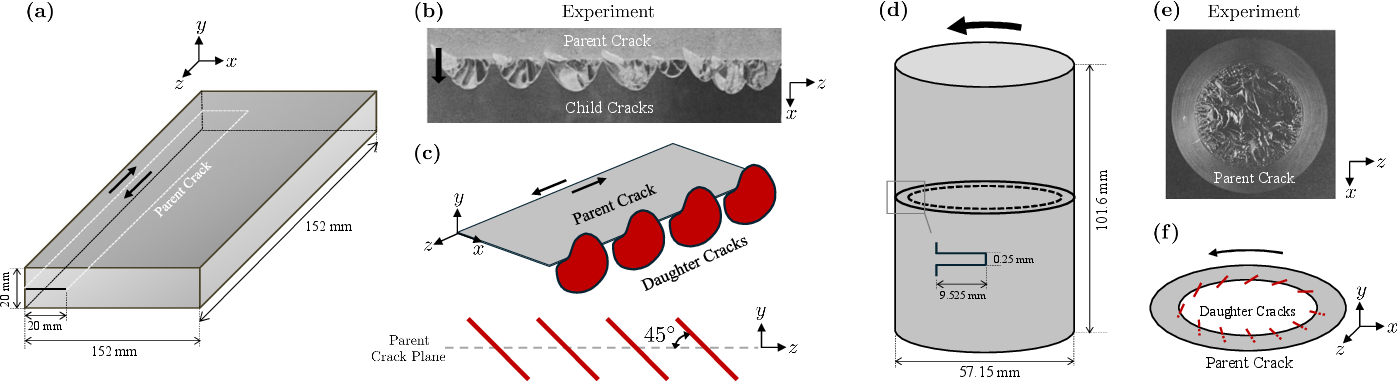}
	\caption{{{ {(a) Schematic of the tearing test (not to scale) over a thick plate with the initial parent crack from Knauss \cite{knauss1970observation}.} (b) Experimental observation of echelon crack growth (top view). The parent crack fragments into multiple { child} cracks. The direction of crack growth is shown with a black arrow. (c) Schematic of the nucleated { child} cracks in a 3D view and a side view. (d) Schematic of the torsion test over a notched cylinder \cite{knauss1970observation}. (e) Echelon crack growth and coalescence originating from the parent crack as observed in the experiment. (f) Schematic of the nucleated { child} cracks aligned at approximately 45 degrees with respect to the parent crack plane.}} }  \label{Fig1}
\end{figure*}

The analysis of the growth of pre-existing large cracks in brittle materials under quasi-static loading is commonly separated into two questions: (i) when a crack grows, and (ii) where it grows. The first question was resolved more than a century ago by Griffith \cite{Griffith21}; along a known path, a crack grows when
\begin{equation}
-\dfrac{\partial{\mathcal{W}}}{\partial \Gamma} = G_c, \label{Griffith}
\end{equation}
where $G_c$ is the fracture toughness, { $\mathcal{W}$ is the total potential energy}, and $\Gamma$ denotes the crack surface area; however, experimental validation, especially in soft elastic brittle materials, was not completed until the 1950s. By contrast, the second question, concerning the crack path and morphology, is considerably more difficult. When a planar crack is subjected to loading that is not purely tensile (mode I), it may deviate from its original plane through either abrupt kinking or gradual curving. Under predominantly anti-plane shear (mode III)  loading, the \emph{parent} crack front often undergoes a sudden fragmentation into multiple, initially disconnected \emph{child} cracks (also known as \emph{daughter} cracks) \cite{sommer1969formation, knauss1970observation, ronsin2014crack, pham2016growth, fineberg2022hidden}. Owing to their characteristic stepped geometry, these fracture patterns are commonly referred to as \emph{echelon} cracks (Fig.~\ref{Fig1}). 

The study of echelon crack formation is of significant theoretical and practical importance.  From a theoretical perspective, it represents a striking instability that arises under an otherwise simple and common mode of loading, revealing unexpectedly complex crack morphogenesis processes.  Remarkably, this   phenomenon appears across all brittle materials, including glass, rocks, polymers, and hydrogels, and over many length scales, which is why it has drawn sustained attention across science and engineering \cite{cooke1996earthquake, zimmermann2009bone, ronsin2014crack, zehnder2015spiral, itamar2018topological, fineberg2022hidden}. From a practical standpoint, echelon cracking is relevant to any system experiencing twisting or shear out-of-plane.

%

One of the earliest experimental investigations of echelon crack growth was performed by Knauss \cite{knauss1970observation}, who conducted tearing tests on thick, notched plates of Solithane polymer. He observed the emergence of { child} cracks inclined at approximately $45^{\circ}$ to the original crack axis; see Fig.~\ref{Fig1}(a-c). He also reported similar crack growth for torsion of an externally notched cylinder (Fig.~\ref{Fig1}(d-f)).  Subsequent studies have reported echelon crack formation in other brittle materials { \cite{cooke1996earthquake, abass1996hydrostone, ronsin2014crack, zehnder2015spiral}.} Their occurrence is known to depend on a combination of geometric and loading conditions, as well as intrinsic material properties. 
Experiments on hydrogels further indicate that echelon crack formation depends on the intrinsic material length scales at the crack tip \cite{ronsin2014crack}.

Existing fracture theories remain unable to account for echelon crack formation, and in particular fail to satisfactorily address two fundamental questions: (i) why does the crack fail to propagate in a self-similar manner along the axis of the pre-existing crack \cite{barenblatt1961brittle}, and (ii) if the crack tends to propagate out of its original plane, why does it do so through fragmentation into multiple cracks rather than via smooth, continuous deviation?

The first question has been examined using the various empirical crack path criteria proposed in the literature since the 1960s \cite{ErdoganSih1963, sih1974strain, goldstein1974brittle, nuismer1975energyrelease, wu1978fracture}, together with the Griffith criteria (\ref{Griffith}).
Broadly, these criteria, developed primarily for linear elastic, isotropic materials, fall into two classes: (i) energy-based criteria, which postulate crack growth along paths of maximum energy release rate \cite{nuismer1975energyrelease}, and (ii) local stress-field-based criteria, which postulate that the crack selects a path that locally restores mode I conditions \cite{lazarus2008comparison}.
For many two-dimensional crack growth problems in compressible isotropic materials, these criteria often lead to similar predictions { \cite{bittencourt1996, bouchard2003, paulino2007}}. However, their extensions to three dimensions, nonlinear elastic or anisotropic materials, are highly nontrivial \cite{chambolle2009}.  
Proposed 3D extensions in the literature indicate that, for the experimental configuration studied by Knauss (Fig.~\ref{Fig1}) and other similar mode III tests, energy-based criteria predict continued planar crack growth \cite{pham2017phase, mittelman2015energy}. In contrast, stress-based criteria such as the Maximum Tangential Stress (MTS) criterion or the Principle of Local Symmetry (PLS) generally predict that the crack will smoothly curve out of the plane \cite{lazarus2008comparison}. Similar theoretical models have seen limited development in soft materials.

To explain why an out-of-plane crack fragments into multiple cracks, several empirical hypotheses have been proposed. One suggests that, under mixed-mode I+III loading, sinusoidal or helicoidal perturbations of the crack front become unstable \cite{karma2010helical, leblond2011theoretical}; however, the associated linear instability criteria lack experimental support \cite{pham2014further}. Moreover, later experiments indicated that fragmentation arises from the nucleation of disjointed { child} cracks, rather than from the continuous evolution of an unstable crack front \cite{ronsin2014crack, pham2016growth, itamar2018topological}. 
Another view holds that a smoothly curving crack cannot maintain locally mode I conditions and therefore must fragment \cite{pham2014further}, although this has not been simulated, and the underlying rationale for local mode I growth and its general applicability are not understood.

Consequently, the inability to predict echelon crack formation reveals a significant gap in the understanding of fracture.
Addressing this gap requires a theoretical and computational framework that can predict this phenomenon, while also providing a general description of crack nucleation and propagation across arbitrary loading and geometrical configurations, rather than being tailored specifically to the echelon crack problem. 

Over the past two decades, the development of the variational theory of brittle fracture and its regularized phase-field formulation \cite{Francfort98, Bourdin00} has raised the prospect of such a unified framework for determining both when and where pre-existing cracks grow. More recently, extensions of this framework that explicitly incorporate material strength have provided a complete description of crack nucleation in the absence of a pre-existing crack singularity, a regime that the purely energetic variational theory cannot capture \cite{KFLP18, KBFLP20}. When applied to the growth of pre-existing cracks under mode I, mixed mode I+II, and even certain mode I+III loading conditions, these two formulations have previously yielded similar predictions \cite{KDK2025Comparison}.  In the present work, to address the two questions posed above regarding echelon crack formation, we confront these models with the classical experiments of Knauss.

\section{Analysis with the classical variational model}

The variational theory of brittle fracture is based on Griffith’s conceptual tenet of viewing the crack's state through a cost-benefit analysis: the crack has a given length at that time because having a longer crack would cost more surface energy than it would save in potential energy.
In this theory, the deformation field $\mathbf{y}(\mathbf{X},t)$ and the crack set $\Gamma(t)$, under quasi-static { displacement-controlled loading}, are obtained by globally minimizing the total energy, defined as the sum of elastic and fracture contributions:
\begin{equation}
 \mathcal{E}(\mathbf{y}, \Gamma) := \int_{\Omega_0 \setminus \Gamma} W(\mathbf{F}) \, \mathrm{d}\mathbf{X} 
+ G_c \, \mathcal{A}(\Gamma), 
\label{Variational} 
\end{equation}
where $\mathcal{A}(\Gamma)$ stands for the surface measure (2–dimensional Hausdorff measure) of the unknown crack and $W(\mathbf{F})$ stands for the hyperelastic energy function of the deformation gradient tensor, $\bfF$. This variational formulation provides a more rigorous description of crack growth without recourse to empirical stress-based criteria. However, several fundamental mathematical questions have remained open. For instance, even when global minimality in the sharp theory is relaxed to a notion of local stability (metastability), analysis show that crack kinking is incompatible with smooth crack evolution \cite{chambolle2009, chambolle2010revisiting, francfort2022twenty}.

Nevertheless, this approach has been highly successful in predicting the growth of large pre-existing cracks across a wide range of linear and nonlinear boundary-value problems.
In numerical simulations, the sharp crack formulation is commonly regularized using a phase-field approximation \cite{Bourdin00}, now known as the classical variational phase-field model. A phase field $v = v (\bfX, t)$ is introduced to regularize the crack surface, which takes values in the range [0, 1] over a phase boundary of infinitesimal width $\varepsilon$. Precisely, $v = 1$ identifies regions of the sound material, whereas $v < 1$ identifies regions of the material that have been fractured. The pair of displacement field and phase field minimizes the functional
\begin{equation}
\mathcal{E}^\eps(\bfy,v)
=\int_{\Omega_0}\!\left[
v^2 {W}(\bfF)
+\frac{3G_c}{8}\!\left(
\frac{1-v}{\eps}
+\eps\,\nabla v\!\cdot\!\nabla v
\right)\right]\mathrm{d}\bfX,
\label{BFM00}
\end{equation}
subject to the irreversibility condition on the phase field.
The regularized functional $\mathcal{E}^{\eps}$  (\ref{BFM00}) $\Gamma$-converges to $\mathcal{E}$ (\ref{Variational}) for $\varepsilon \rightarrow 0$. 
{ The particular choice of regularization in (\ref{BFM00}) is referred to as the AT1 model in reference to the work of Ambrosio and Tortorelli \cite{ambrosio1990}. Other regularizations of (\ref{Variational}) are possible; however, the AT1 model is often preferred because it is the simplest formulation that yields a physically meaningful phase-field evolution with compact support. }
The success of this formulation for a wide range of problems has fostered the view that the variational framework largely resolves the classical problem of the growth of large cracks in brittle materials initiated by Griffith.

\begin{figure*}[h!]
	\centering
	\includegraphics[width=0.8\textwidth]{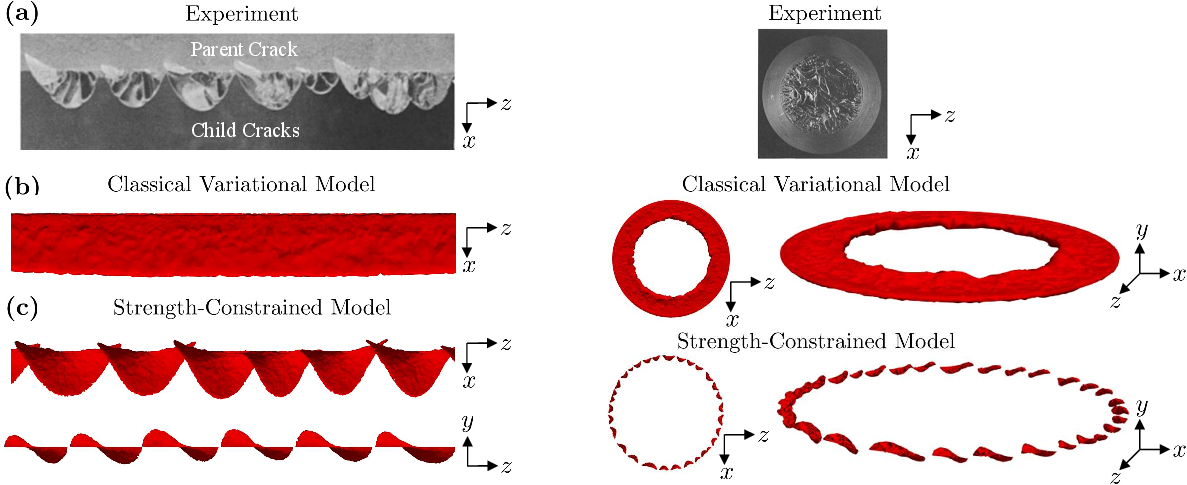}
	\caption{{Comparison of fracture model predictions with tearing experiment in a thick notched plate (left) and torsion experiment in a notched circular cylinder. (a) Experimental results from \cite{knauss1970observation}. (b) Planar crack path predicted by classical variational phase-field model (\ref{BFM00}). (c) Echelon crack path predicted by the strength-constrained phase-field model (\ref{phase-field-equations}). Only the { child} cracks are shown from two views.} }  \label{Fig1-num}
\end{figure*}

\begin{figure*}[h!]
	\centering
	\includegraphics[width=0.9\textwidth]{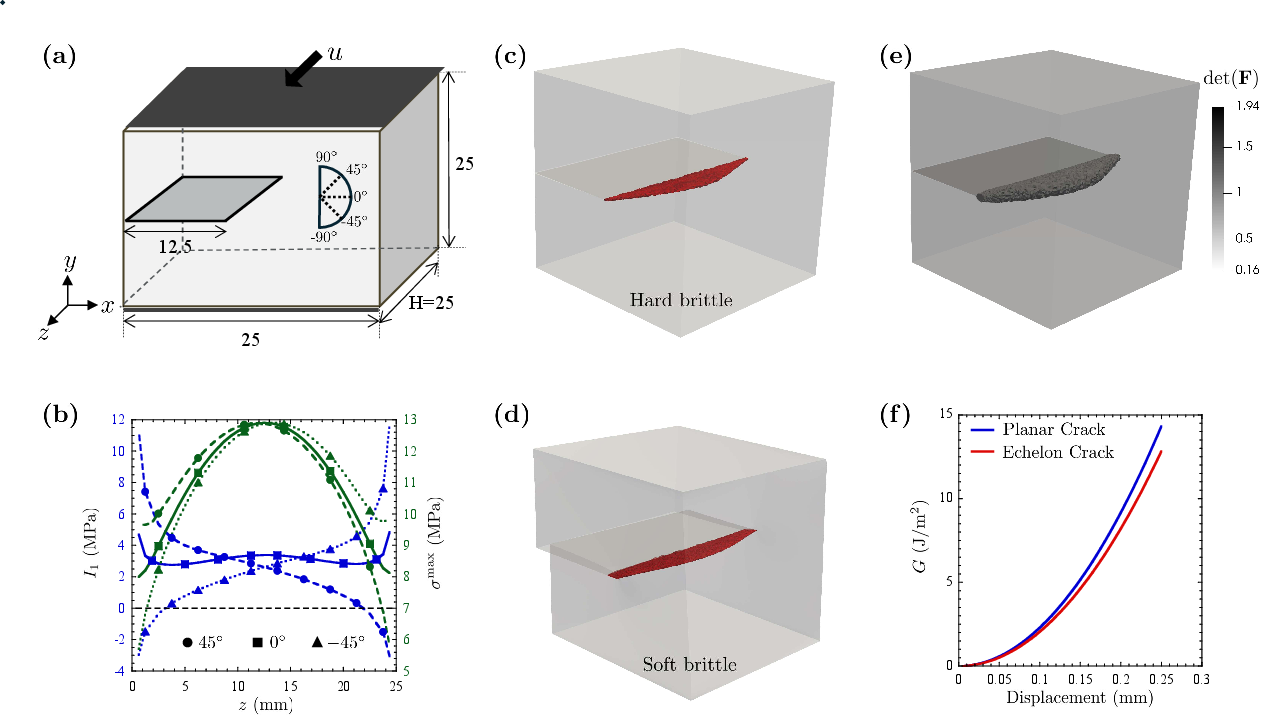}
	\caption{{(a) Schematic of a smaller geometry used for comprehensive analysis (all dimensions are in mm; applied displacement in $z$-direction). The angular positions shown in the range [-90$^\circ$, 90$^\circ$] correspond to positions along a circular arc of fixed radius 0.02 mm ahead of the crack front, where stresses are evaluated. (b) Plots of the first invariant of the stress tensor, $I_1$ (left axis) and maximum principal stress, $\sigma_{\rm max}$ (right axis), evaluated at a fixed radial distance ahead of the crack for angular coordinates -45$^\circ$, 0$^\circ$, and 45$^\circ$, and shown as functions of the position $z$ along the crack front.
    (c) Crack path contour obtained from the variational model for a hard brittle material, Graphite, and (d) contour for soft brittle material, PDMS. (e) Plot of the determinant of the deformation gradient tensor over the propagating crack. (f) Plot of the energy release rate computed using (\ref{energy-release-rate}) for planar and echelon crack growth.}}  \label{Fig2}
\end{figure*}

However, when we applied the variational phase-field model to simulate Knauss’s tearing and torsion experiments, echelon crack growth is not reproduced; instead, the model predicts continued planar crack propagation, as shown in Fig.~\ref{Fig1-num}(b). Owing to the substantial computational cost associated with the three-dimensional experimental-scale geometry \footnote{{ Leveraging adaptive mesh refinement algorithms can significantly mitigate the computational expense of three-dimensional phase-field fracture simulations, as demonstrated recently in \cite{gupta2024adaptive, nguyen2025adaptive}.  }}, we focus on the tearing test and adopt a reduced geometry for further analysis (Fig.~\ref{Fig2}(a)), which enables a more detailed study.
The analysis is conducted for two isotropic brittle materials representing hard and soft responses: graphite, modeled as a linear elastic material, and PDMS, modeled as a nearly incompressible nonlinear elastic material. Constitutive behavior for the linear and nonlinear elastic materials is discussed in the Appendix A. Material parameters for graphite are taken from \cite{sato1987graphite, KBFLP20}, and those for PDMS from \cite{Poulain17, KKLP24}, also included in Appendix A.

The results obtained with the reduced geometry for graphite and PDMS are shown in Fig.~\ref{Fig2}(c–d). In both cases, the model again predicts planar crack growth.
The variational phase-field framework has previously been applied to study echelon cracking in hard brittle materials, with similar outcomes reported \cite{pham2017phase, molnar2024phase}. However, in those studies, it was argued that echelon crack growth can be recovered by introducing material disorder---through defects or stochastic fracture toughness---and by modifying the variational formulation via a spectral strain-energy decomposition. They modified the energy functional in the classical variational formulation in the following manner:
\begin{equation}
\begin{aligned}
\hat{\mathcal{E}}^\eps(\bfy,v)
&=\int_{\Omega_0}\!\left[v^2 W^{+}(\bfF)+W^{-}(\bfF)\right]\mathrm{d}\bfX
+\frac{3G_c}{8}
\int_{\Omega_0}\!\left(
\frac{1-v}{\eps}
+\eps\,\nabla v\!\cdot\!\nabla v
\right)\mathrm{d}\bfX,
\end{aligned}
\label{BFM00-split}
\end{equation}
where $W^{+}$ denotes the tensile part of the strain energy degraded by the phase field and $W^{-}$ the compressive part that does not drive fracture. The motivation for this decomposition is that the standard regularized formulation (\ref{BFM00}) does not distinguish between crack faces under tension and compression: in the sharp crack setting, compressive loading enforces contact and allows force transmission, whereas in the regularized model, all energy is degraded, preventing such transfer. 
This leads to violations of material impenetrability and permits crack growth under compression, contrary to experimental observations.
Although the split formulation (\ref{BFM00-split}) has been proposed as a remedy, there is no rigorous and general procedure for decomposing an arbitrary strain-energy density into tensile and compressive parts in two or three dimensions \cite{KDK2025Comparison}. In particular, the spectral decomposition used in prior work \cite{pham2017phase, molnar2024phase} { based on the fully variational formulation introduced by Miehe et al. \cite{miehe2010}  } introduces unphysical residual shear stresses in cracked regions \cite{LK24}. { The corresponding elastic energy is removed from the crack driving force,  rendering it unphysical for shear-dominated fracture. For the tearing problem, this would mean that the energetic advantage of planar crack growth is artificially reduced, making it more likely for the echelon crack growth to traverse the energetic advantage if some out-of-plane material disorder is introduced.}

Furthermore, our analysis shows that an energy decomposition should not be needed for the present problem. As shown in Fig.~\ref{Fig2}(e), even in the large-deformation regime, the determinant of the deformation gradient remains well above zero, indicating that material interpenetration does not occur and crack faces are not in contact; for small deformations, ${\rm det}(\mathbf{F})$ remains close to unity. Stress analysis further supports this conclusion. 
{  Stresses are evaluated on a circular arc of radius 0.02 mm in the local plane normal to the crack front, with the arc centered at the crack front. For each location $z$ along the crack front, stresses are sampled at angular positions \{-90$^\circ$, -45$^\circ$, 0$^\circ$, 45$^\circ$, 90$^\circ$\}, as illustrated in Fig.~\ref{Fig2}(a). }
For planar crack growth (corresponding to the $0^\circ$ orientation in Fig.~\ref{Fig2}(a)), the first stress invariant $I_1$ is strictly positive ahead of the crack across the thickness (Fig.~\ref{Fig2}(b)), indicating a substantial opening mode; the normal stress component $S_{yy}$ is also positive.
As an additional check, we computed the energy release rate, $G$, for both planar and echelon cracks using
\begin{equation}
G \approx -\frac{\mathcal{W}(u, A + \Delta A) - \mathcal{W}(u, A)}{\Delta A},
\label{energy-release-rate}
\end{equation}
where $\mathcal{W}= \int_{\Omega_0} W(\bfF) \,{\rm d}\bfX$ is the total strain energy, $A$ is the initial crack surface area, and $\Delta A = A/50$ is the area increment. { The planar crack was generated geometrically, whereas the echelon crack geometry was generated with the strength-constrained phase-field model described in the next section. 
Specifically, a simulation with the strength-constrained phase-field model was first performed to generate the echelon crack configuration corresponding to $\Delta A = A/50$. An elastic analysis was then carried out on the resulting cracked geometry to compute $\mathcal{W}(u, A+\Delta A)$.} The results, shown in Fig.~\ref{Fig2}(f), demonstrate that the planar crack has a higher energy release rate and is therefore energetically favored, consistent with the phase-field predictions. { We note here the previous work of Mittelman and Yosibash \cite{mittelman2015energy}, who conducted a more comprehensive energy release rate analysis by introducing idealized geometrical echelon cracks, and reached the same conclusion.}

We therefore conclude that the variational formulation, rooted in energy competition, is insufficient to predict echelon crack growth. Thus, contrary to the widespread belief in the literature, it should be considered fundamentally incomplete for predicting large crack growth under tensile/shear regimes.

\section{Analysis with the strength-constrained phase field model}

{ Classical variational phase-field model (\ref{BFM00}) lack a critical capability: the ability to describe crack nucleation in specimens without a pre-existing crack} \footnote{ { For finite values of regularization length $\varepsilon$, the uniaxial tensile strength of the material can be incorporated into the classical variational phase-field model at the cost of severing connection with the sharp fracture model (\ref{Variational}). However, the resultant strength surface only agrees with the physical strength surface at the uniaxial tensile stress point in the stress space.}} This limitation arises from the absence of a strength criterion. While strength has been previously deemed unnecessary for analyzing the growth of large pre-existing cracks in the literature, recent experiments have cast echelon crack formation as a nucleation problem rather than a continuous evolution of the parent crack front \cite{pham2016growth}. Motivated by these experiments,
we investigate the tearing problem using a strength-constrained phase-field formulation. Such a formulation has been developed in recent years \cite{KFLP18, KBFLP20} for nucleation of cracks as a generalization of the classical variational model. It has been widely validated across a range of problems \cite{KLP21, KRLP22, KLDLP23, LK24, KKLP24, kamarei2026nine} and has been shown to accurately predict arbitrary crack nucleation. Notably, the formulation has been validated recently for a range of mixed-mode fracture problems, especially those involving mode I+II loading \cite{KDK2025Comparison}. 

In this formulation, the material strength is represented as a surface in three-dimensional stress space, known as the strength surface. This surface defines the set of critical stress states $\mathbf{S}$ at which the material fractures under monotonically increasing, spatially uniform, but otherwise arbitrary loading. { The surface is represented as}
\begin{equation}
\mathcal{F}(\bfS)=0. \label{Strength surface}
\end{equation}
For isotropic materials, a simple choice is the two-parameter Drucker–Prager (DP) strength surface, which has been shown to capture the fracture strength of many nominally brittle materials. { We adopt this form in the present work. More general strength surfaces, including those with explicit dependence on the Lode angle, may nevertheless be required to accurately reproduce the experimentally measured strength behavior of certain materials \cite{chockalingam2025MCHB}.}

The strength surface violation acts as a necessary but not sufficient condition for crack evolution \cite{LK24}. It acts as a constraint on the variational formulation. { Mathematically, we can state it as the following \cite{lopez2025Whenandwhere}: at a discrete time $t_k$ the pair of deformation field and the crack set  $\left( \bfy_k := \bfy_k(\mathbf{X}, t_k), \; \Gamma_k := \Gamma(t_k) \right)$ minimize the functional (\ref{Variational}) among all ($\bfy, \Gamma= \Gamma_{k-1} \cup \Delta \Gamma$) with
\begin{equation}
\Delta \Gamma \subset \mathcal{V}_{\mathcal{F}}(t) \quad {\rm and} \quad \Gamma \supset \Gamma_{k-1},
\end{equation}
where $\mathcal{V}_{\mathcal{F}}(t) = \left\{ \mathbf{X} : \mathcal{F}(\mathbf{S}(\mathbf{X},t)) \ge 0 \right\}$.} The phase-field regularization of this model is detailed in the Appendix B; a finite-element implementation of the model in the open-source platform FEniCS to solve the echelon crack problem is available on GitHub \cite{KumarEchelonCode2026}.

We applied the strength-constrained phase-field model to the full experimental geometry of Knauss for the tearing and torsion test (Fig.~\ref{Fig1}(a),(d)), without introducing stochastic material properties or a defect distribution around the crack. The initial crack surface is planar, with no imposed crack-front undulations. 
Remarkably, the model naturally predicts the formation of echelon cracks for both tests, as shown in Fig.~\ref{Fig1-num}(c). 
The cracks are inclined at an angle of approximately $40^{\circ}$ to the original crack axis, which is close to the experimentally reported value. The critical load was not reported from the experiments, making more quantitative comparisons difficult. As discussed later, a complete validation requires independent measurement of the shear strength, which has not been performed in previous work.

Importantly, this is the same phase-field model previously validated across a wide range of challenging problems, such as indentation, trousers, and pokerchip tests \cite{KLP21, KRLP22, kamarei2026nine}, without any problem-specific modifications. The material parameters in the model---elasticity, toughness, and strength---are standard, experimentally measurable quantities as discussed in Appendix A. Thus, unlike prior analyses, the strength-constrained model predicts echelon crack growth without ad hoc assumptions.

For further study, we again adopt the smaller geometry (Fig.~\ref{Fig2}(a)) and simulate the problem for graphite and PDMS, capturing the fragmentation and evolution of the initial crack front. See Appendix C for a discussion of the formation and growth of echelon cracks; see also supplementary material, movies S1–S4.

\begin{figure*}[h]
	\centering
	\includegraphics[width=0.8\textwidth]{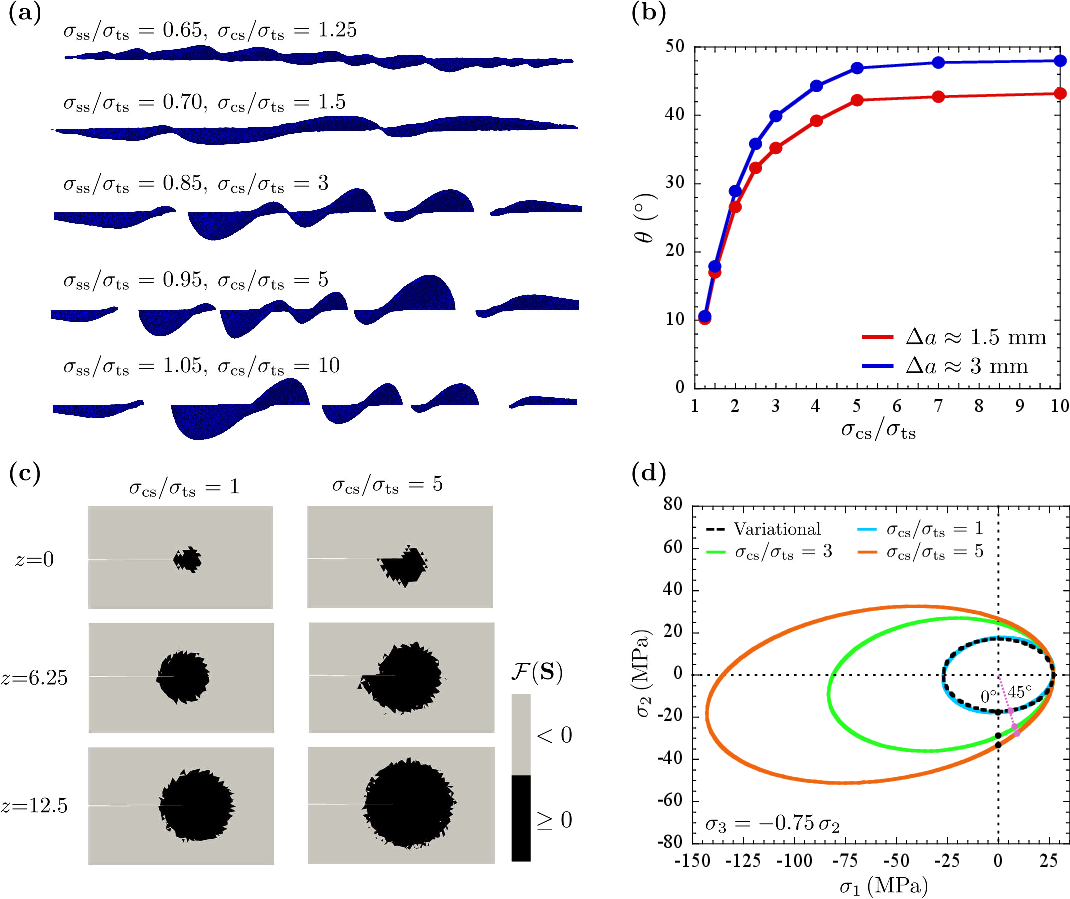}
	\caption{(a) Crack contours for five increasing values of the shear-to-tensile strength ratio, $\sss/\sts$, along with the corresponding compressive-to-tensile strength ratios, $\scs/\sts$. (b) Orientation angle for the largest { child} crack as a function of $\scs/\sts$ for two values of crack extension $\Delta a$. (c) Contour plot around the crack front of the regions of the specimen where the stress field exceeds the strength surface ($\mathcal{F}(\bfS)=0$) at three different locations (coordinate in mm) along the thickness and applied displacement $u=0.075$ mm. (d) { A 2D cut of the strength surface for different $\scs/\sts$ ratios are plotted in terms of the principal stresses $\sigma_1$ and $\sigma_2$ corresponding to the $\sigma_3 = -0.75 \, \sigma_2$ plane. The smallest contour corresponds to $\scs/\sts=1$, the intermediate contour to $\scs/\sts=3$, and the largest contour to $\scs/\sts=5$. Strength surface corresponding to the classical variational model is shown through a dotted line.}}\label{Fig4}
\end{figure*}

The success of the strength-constrained phase field model over the classical variational model highlights the importance of accounting for the strength surface.
To understand it further, we examine the influence of a key non-dimensional parameter---the ratio of shear to tensile strength, $\sss / \sts$, which directly enters the model. For graphite, we vary this ratio from 0.65 to 1.05. In terms of the corresponding ratio of compressive to tensile strength,{ $\scs / \sts = \sqrt{3}\, \sss  / (2 \, \sts - \sqrt{3}\, \sss)$}, this spans 1.25 to 10. Crack contours for a maximum extension of $\Delta a = 3$ mm are shown in Fig.~\ref{Fig4}(a).

We observe that for large strength ratios, echelon crack growth occurs, whereas for small ratios the phase-field crack front remains connected and undulating, with the amplitude of the undulations diminishing as $\sss / \sts$ approaches 0.65, ultimately yielding an almost planar front. The orientation angle of the largest crack, $\theta$, was computed for two crack extensions, $\Delta a=$1.5 mm and 3 mm, in each case and plotted in Fig.~\ref{Fig4}(b). The angle transitions from approximately $45^\circ$ for large $\sss / \sts$ (or $\scs / \sts$) to near $0^\circ$ for small values. The angle evolves as the crack gets longer. Since most brittle materials have $\scs / \sts > 3$, an angle of roughly $45^\circ$ for large ratios aligns with experimental observations and the largest tensile stress orientation under pure mode III, whereas maximum energy release rate criteria predict planar growth ($\theta = 0^\circ$).

{This analysis demonstrates that the strength-constrained model interpolates naturally between energy-based and stress-based predictions depending on the shear-to-tensile strength ratio. Consequently, the different empirical stress and energy-based criteria proposed in the literature can be reconciled within this unified framework.}

To further understand why crack orientation depends on $\sss / \sts$ or $\scs / \sts$, we plot a two-dimensional cut of the strength surface in terms of the principal stresses $\sigma_1$ and $\sigma_2$ for different values of $\scs / \sts$ in Fig.~\ref{Fig4}(d). 
For comparison, we also include a strength surface inferred from the classical variational model, even though it is not explicitly included in that formulation. The 2D cut is taken along the plane $\sigma_3 = -0.75 \, \sigma_2$, which approximates the relationship between the maximum and minimum principal stresses ahead of the crack in this tearing problem. For $\scs / \sts = 1$, the strength surface closely matches the inferred surface from the variational model, explaining why both models predict similar crack growth under this condition.

We also examine the directions of stress evolution in front of cracks oriented at $0^\circ$ and $45^\circ$. The analysis in Fig.~\ref{Fig4}(c) shows that violations of the strength surface occur more readily at $45^\circ$ for larger $\scs / \sts$. The strength constraint enforces that the growth of a large crack occurs only in regions where the strength criterion is exceeded. Hence, when shear strength is high, the material preferentially fails under tension, causing the parent crack to curve out of the plane and often fragment into { child} cracks. Fig.~\ref{Fig4}(c) shows the regions near the crack front where the strength surface is exceeded for an applied displacement of $u=0.075$ mm. As one moves away from the mid-plane of the domain ($z=12.5$ mm), the strength violation becomes increasingly asymmetric for larger values of $\scs/\sts$.
Conversely, when shear strength is low, the crack favors planar propagation under shear since the strength violation is symmetric around the crack front across the thickness (Fig.~\ref{Fig4}(c)). While this behavior is intuitive, crack formation is also influenced by the material’s fracture toughness. In regions exceeding the strength surface, the strength-constrained model drives crack growth by minimizing the combined elastic and surface energies \cite{lopez2025Whenandwhere}. Therefore, the precise pattern of crack formation cannot be predicted by strength considerations alone.

\begin{figure}[h!]
	\centering
	\includegraphics[width=4.in]{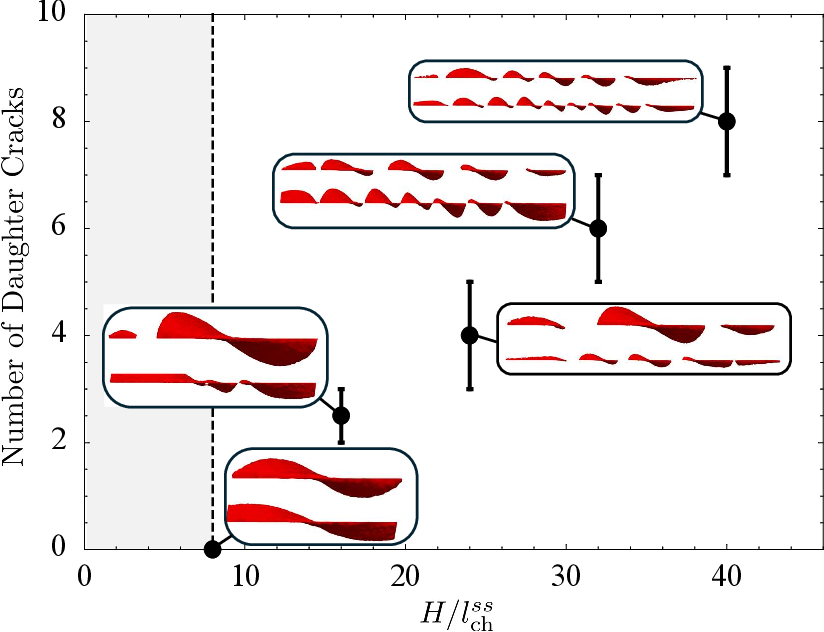}
	\caption{{ Number of { child} cracks as a function of $H/ l_{\rm ch}^{ss}$, where $H$ is the plate width and $l_{\rm ch}^{ss}$ is the shear-based fracture length scale (\ref{length-scale}). Inset panels (top) show results obtained by varying $l_{\rm ch}^{ss}$ via changes in $G_c$, $E$, or $\sss$; inset panels (bottom) show results obtained by varying $H$. Error bars indicate the range of observed crack counts. No echelon cracks observed in the gray shaded region.}}\label{Fig5}
\end{figure}

We further investigated the influence of material toughness and, more generally, the fracture process zone size on echelon crack formation. The fracture process zone is associated with a characteristic length scale emerging from the governing equations (\ref{phase-field-equations}). Specifically, the elastic energy density $W(\mathbf{F})$ and the strength surface $\mathcal{F}(\mathbf{S})$ both have dimensions of force per unit area, while the toughness $G_c$ has dimensions of force per unit length. Their combination defines a family of intrinsic material length scales. For the present problem, where the stress state is predominantly shear, we consider the linear-elastic shear-based length scale 
\begin{equation}
l_{\rm ch}^{ss}=\frac{3 G_c E}{8 \sss^2 },
\label{length-scale}
\end{equation}
where $E$ is the Young's modulus. { We performed three sets of simulations, individually varying $G_c$, $E$, and $\sss^2$ by factors ranging from 1 to 5 from their baseline values listed in Table 1 (Appendix A) while keeping all other material and geometric parameters fixed. The corresponding ranges were $G_c \in [91, 455]$ J/m$^2$, $E \in [9.8, 49]$ GPa, and $\sss \in [10.3, 23]$ MPa.}  For each case, we recorded the number of { child} cracks that formed. The regularization length $\varepsilon$ was maintained at $\varepsilon / l_{\rm ch}^{ss} = 1/3$. We observe that the initial number of cracks increases as $\varepsilon / l_{\rm ch}^{ss}$ decreases, although it converges for sufficiently small values. Once cracks have grown to approximately $10 \, \varepsilon$, the number of { child} cracks is largely insensitive to $\varepsilon / l_{\rm ch}^{ss}$, as smaller cracks either coalesce rapidly or stop growing due to shielding effects.

Remarkably, both the number and orientation of { child} cracks were found to depend primarily on the value of $l_{\rm ch}^{ss}$, regardless of whether $G_c$, $E$, or $\sss$ was varied. The results, shown in Fig.~\ref{Fig5} (top row of each inset panel), indicate that the number of { child} cracks decreases as $l_{\rm ch}^{ss}$ increases. For sufficiently large $l_{\rm ch}^{ss}$, echelon cracks no longer form, and the crack evolves as a continuous, curved front. These observations are consistent with Ronsin et al. \cite{ronsin2014crack}, who reported that dissipative gels under a lower energy release rate, associated with a smaller process zone \cite{slootman2020creton}, exhibit echelon cracking, whereas under a higher energy release rate, favor smooth, curved cracks.

We also performed simulations keeping $l_{\rm ch}^{ss}$ constant while varying the plate width $H$ (Fig.~\ref{Fig2}(a)) by a factor of 1–5, maintaining $\varepsilon / l_{\rm ch}^{ss}$ fixed. The results, shown in Fig.~\ref{Fig5} (bottom row of each inset panel), reveal that reducing $H$ by a factor of 5 has a similar effect as increasing $l_{\rm ch}^{ss}$ fivefold. This suggests that echelon crack formation is governed by the non-dimensional parameter $H / l_{\rm ch}^{ss}$.

\section{Conclusions}

We have shown through the analysis of a classical experiment that Griffith's theory, when cast into the variational phase-field model, is incomplete for predicting the growth of a large crack. The underlying reason is that the energetic Griffith theory does not account for the restrictions imposed by the strength surface on crack evolution. 

A phase-field model that performs strength-constrained minimization naturally reproduces echelon crack formation and evolution without defects, stochasticity, or ad hoc assumptions. { While material disorder is present in real materials and may affect crack morphology, the results in this work provide evidence that they are not the { sole} cause of echelon crack nucleation. }
Crack orientation and fragmentation are controlled by two key non-dimensional parameters: the shear-to-tensile strength ratio, $\sss / \sts$, and the plate-thickness-to-characteristic-length ratio, $H / l_{\rm ch}^{ss}$ if $H / l_{\rm ch}^{ss}$ is sufficiently large. Systematic experimental measurements and control of shear strength and the $\sss / \sts$ ratio are needed to fully validate the results of this work. Unnotched torsion test provides a way to independently measure shear strength as discussed in Appendix C with some additional results.
{While the present study employs a Drucker–Prager strength surface, alternative strength surfaces \cite{chockalingam2025MCHB} may be adopted as more comprehensive strength data become available. Nevertheless, the general conclusions of this work are expected to be robust with respect to the particular choice of strength surface.}
Furthermore, additional experimental configurations, including those with a small mode III component \cite{pham2014further, fineberg2022hidden}, will provide natural extensions for applying and testing the predictive scope of the present framework. 

The strength-constrained formulation is similar to the classical variational formulation of fracture, differing only through the introduction of an additional constraint. Accordingly, its analytical treatment is expected to follow many of the same lines as that of the unconstrained formulation. It is also likely to inherit several of the well-known mathematical challenges of the classical theory, such as the rigorous understanding of crack kinking, and may also introduce new difficulties. A comprehensive mathematical analysis of the sharp-interface limit, namely the phase-field equations as the regularization length $\varepsilon \rightarrow 0$, remains to be carried out.

Despite these open problems and the need for more experimental validation, when combined with validation studies from previous works on crack growth under mode I and mode I+II loading, the results presented in this work demonstrate that the numerical strength-constrained phase-field framework \cite{KFLP18, KBFLP20} can robustly predict arbitrary fracture initiation and propagation in both soft and hard isotropic brittle materials. 

 \section*{Acknowledgements}

\noindent The authors would like to acknowledge the financial support from the National Science Foundation, United States, through the grant CMMI-2404808.

\section*{Appendix A. Constitutive modeling of graphite and PDMS}

\noindent \textbf{Constitutive behavior of hard brittle materials.} For linear elastic isotropic materials such as graphite, the strain energy density function is written as
\begin{equation}
	{W}(\bfE(\bfu)) =\frac{E}{2(1+\nu)} \, {\rm tr}\,\bfE^2+\dfrac{E\nu}{2 (1+\nu)(1-2\nu)}({\rm tr}\,\bfE)^2,\label{W-mu}
\end{equation}
where $E$ is the Young's modulus, $\nu$ is the Poisson's ratio, and $\bfE(\bfu)$ is the infinitesimal strain tensor
\begin{equation*}
\bfE(\bfu)=\dfrac{1}{2}(\bfF+\bfF^T - 2 \bfI),
\end{equation*}
with $\bfF$ being the deformation gradient tensor. The stress tensor at any material point $\bfX$ and time $t\in[0,T]$ is given by
\begin{equation*}
	\bfS(\bfX,t)=\dfrac{\partial {W}}{\partial \bfE}(\bfE).
\end{equation*}
The elastic constants can be measured with uniaxial tension tests. A number of standard tests, such as the Compact Tension test, also exist for measuring fracture toughness.  The strength surface measurement can be conducted by carrying out experiments on thin tubes subjected to a combination of axial force and inner pressure \cite{KBFLP20}. Such experiments were performed by Sato \cite{sato1987graphite} on graphite. The reported values are listed in the table below. \\

\begin{table}[htbp]
\centering
\caption{Mechanical properties of graphite.}
\label{tab:graphite}
\begin{tabular}{lcc}
\hline
Property & Symbol & Value \\
\hline
Young's modulus        & $E$                     & 9.8 GPa \\
Poisson's ratio        & $\nu$                   & 0.13 \\
Fracture toughness     & $G_c$                   & 91 J/m$^2$ \\
Tensile strength       & $\sigma_{\mathrm{ts}}$  & 27 MPa \\
Shear strength         & $\sigma_{\mathrm{ss}}$  & 23 MPa \\
\hline
\end{tabular}
\end{table}

\noindent \textbf{Constitutive behavior of soft brittle materials.}
Nonlinear elastic materials such as PDMS are nearly incompressible and show a strain stiffening behavior. A non-Gaussian strain energy function, such as the Lopez-Pamies function \cite{LP10}, can be used to model their elastic behavior:
\begin{equation}
\begin{aligned}
    {W}(\bfF)=&\sum_{r=1}^{2} \frac{3^{1-\alpha_r}}{2\alpha_r} \mu_r \left[ (\mathbf{F} \cdot \mathbf{F})^{\alpha_r} - 3^{\alpha_r} \right] 
    - \sum_{r=1}^{2} \mu_r \ln (\det \mathbf{F}) + \frac{\kappa}{2} (\det \mathbf{F} - 1)^2,
\end{aligned}
\label{eq:energy_function}
\end{equation}
where, $\mu_1$ and $\mu_2$ are shear modulus parameters, such as total shear modulus $\mu= \mu_1 +\mu_2$, $\kappa$ is the bulk modulus, and $\alpha_1$, $\alpha_2$ are strain stiffening parameters. The first Piola-Kirchhoff stress at any material point $\bfX$ and time $t\in[0,T]$ is given by
\begin{equation}
{\bfS}^{(1)}(\bfX,t)=\dfrac{\partial {W}}{\partial \bfF}(\bfF).
\end{equation}
We define the strength surface in terms of the Biot stress tensor $\bfS=({\bfS^{(1)}}^T\bfR+\bfR^T\bfS^{(1)})/2$, where $\bfR$ is the rigid rotation tensor defined through a polar decomposition of the deformation gradient $\bfF=\bfR\bfU$, with $\bfU$ being the right stretch tensor \cite{KKLP24}.

Fracture toughness can be measured directly from a pure shear test, a single-edge notch test, or a tearing test \cite{RivlinThomas53}. The tensile strength can be measured from a uniaxial tensile test. At least one additional strength measurement is needed to construct the Drucker-Prager approximation of the strength surface. Shear strength can be measured from the shear rheology test, but it is challenging to achieve uniform stress due to the material's high compliance. An alternative is to measure the hydrostatic strength from the poker chip test or the Gent-Park test \cite{GentLindley1959, Poulain17, KLP21}. The relationship between hydrostatic strength and tensile and shear strengths for a Drucker-Prager surface is given by the relation (\ref{shs-sts-sss}).
We adopt the elastic properties from \cite{Poulain17} and approximate fracture properties used in \cite{KKLP24}, and list them in table 2 below.

\begin{table}[htbp]
\centering
\caption{Mechanical properties of PDMS.}
\label{tab:pdms}
\begin{tabular}{lcc}
\hline
Property & Symbol & Value \\
\hline
Modulus parameter      & $\mu_1$              & 0.42 MPa \\
Modulus parameter      & $\mu_2$              & 0.07 MPa \\
Stiffening parameter   & $\alpha_1$           & 0.03 \\
Stiffening parameter   & $\alpha_2$           & 7.2 \\
Bulk modulus           & $\kappa$             & 50 MPa \\
Fracture toughness     & $G_c$                & 10 J/m$^2$ \\
Tensile strength       & $\sigma_{\mathrm{ts}}$ & 0.1 MPa \\
Hydrostatic strength   & $\sigma_{\mathrm{hs}}$ & 0.125 MPa \\
\hline
\end{tabular}
\end{table}

We note that while it is common to report a critical stretch or energy from tensile testing of elastomers in place of the critical stress, a unified theory, in general, requires a stress threshold. A simple example demonstrating this fact is the case of an incompressible ball under hydrostatic loading, which will be predicted to never fail if a critical stretch or energy threshold criterion for nucleation is utilized.
The strength surface, $\mathcal{F}(\boldsymbol{\sigma})=0$, cannot generally be reformulated in terms of energy or stretch.

\section*{Appendix B. Phase-field regularization of strength-constrained model}

The strength surface (\ref{Strength surface}) constraint is applied to the variational phase-field model (\ref{BFM00}) through a two-step procedure: (i) consider the Euler-Lagrange equations of the variational principle (\ref{BFM00}) as the primal model, and (ii) introduce the strength surface by adding a stress-based driving force to the Euler–Lagrange equation for the phase-field evolution. The resulting formulation says that, subject to the appropriate initial and boundary conditions, and in absence of body forces and inertia, the deformation field $\bfy$ and the phase field $v$ are obtained from solving two partial differential equations
\begin{equation}
\left\{\begin{array}{ll}
\hspace{-0.15cm} {\rm Div} \left({v}^2 \dfrac{\partial {W}}{\partial \bfF}(\nabla \bfy) \right)=0 ,
\vspace{0.1cm}\\
\hspace{-0.15cm}
\dfrac{3}{4} {\rm Div}\left[\varepsilon \, {\delta^\varepsilon}   \, G_c   \nabla v\right]=2 \, v \, {W}(\nabla\bfy)  +c_\texttt{e} -\dfrac{3}{8} \dfrac{ {\delta^\varepsilon}  \, G_c }{\varepsilon},
\end{array}\right. \label{phase-field-equations}
\end{equation}
subject to $\dot{v}< 0$, where the term $c_\texttt{e}(\bfX,t)$ is the additional driving force and $\delta^\varepsilon$  is a non-negative coefficient, both of whose prescriptions depend on the particular form of strength surface. As explained in a recent work \cite{lopez2025Whenandwhere}, these equations numerically describe a constrained minimization problem: crack growth happens in regions where the strength surface has been met through a minimization of the sum of elastic and surface energies. The strength-constraint enters the formulation due to a characteristic of the above PDE system: the phase field can not evolve from its initial value of $v=1$ unless
\begin{equation}
    2 \, v \, {W}(\nabla\bfy)  +c_\texttt{e} -\dfrac{3}{8} \dfrac{ {\delta^\varepsilon}  \, G_c }{\varepsilon} \geq 0
\end{equation}
Thus, for a suitably constructed $c_\texttt{e}(\bfX,t)$, one can ensure that the phase field only evolves when the strength surface is met. It should be noted, though, that a rigorous proof establishing that the limit 
$\varepsilon \rightarrow 0$ of the PDE system \eqref{phase-field-equations} yields a strength-constrained energy minimization formulation is still lacking.

The formulation allows for an arbitrary choice of the strength surface. In this work, we have adopted the Drucker-Prager (DP) surface, expressed as follows:
\begin{equation}
	\mathcal{F}(\bfS)=\sqrt{J_2}+\gamma_1 I_1+\gamma_0=0 ,\label{DP}
\end{equation}
with
\begin{equation}
	\gamma_0=-\sss, \qquad  \gamma_1=\dfrac{\sqrt{3} \sss-\sts}
{\sqrt{3}\sts},
\end{equation}
where
\begin{equation}
	I_1 = \mathrm{tr}\,\bfS, \quad 
J_2 = \frac{1}{2}\mathrm{tr}\,\bfS_D^2, \quad
\bfS_D = \bfS - \frac{1}{3} (\mathrm{tr}\,\bfS) \bfI ,\label{T-invariants}
\end{equation}
stand for two of the standard invariants of the stress tensor $\bfS$, while the constants $\sts>0$ and $\sss>0$ denote the uniaxial tensile and shear strengths of the material, respectively. Corresponding to the DP surface, 
the functional form for $c_\texttt{e}(\bfX,t)$ was presented recently in \cite{KKLP24}:
\begin{equation}
c_{\texttt{e}}(\bfX,t)
=\beta_2^\eps v^2\sqrt{J_2}
+\beta_1^\eps v^2 I_1
-v\!\left(1-\frac{|I_1|}{I_1}\right)W(\bfF).
\label{cehat}
\end{equation}
where $\beta_1^\varepsilon$ and $\beta_2^\varepsilon$ are $\varepsilon$-dependent coefficients
\begin{equation}
\left\{\begin{array}{l}
\beta^\varepsilon_1=\dfrac{1}{\shs}\delta^\varepsilon\dfrac{G_c}{8\varepsilon}-\dfrac{2 {W}_{\texttt{hs}}}{3\shs}\vspace{0.2cm}\\
\beta^\varepsilon_2=\dfrac{\sqrt{3}(3\shs-\sts)}{\shs\sts}\delta^\varepsilon\dfrac{G_c}{8\varepsilon}+
\dfrac{2{W}_{\texttt{hs}}}{\sqrt{3}\shs}-\dfrac{2\sqrt{3}{W}_{\texttt{ts}}}{\sts}\end{array}\right. . \label{betas}
\end{equation}
Here, ${W}_{\texttt{ts}}$ and ${W}_{\texttt{hs}}$ stand for the values of the strain energy function along uniform uniaxial tension and hydrostatic stress states at which the strength surface is violated. $\shs$ is the hydrostatic strength which for the DP surface is related to shear and tensile strengths through the relation
\begin{equation}
    \shs= \dfrac{\sss \sts}{3 \sss - \sqrt{3} \sts}.
    \label{shs-sts-sss}
\end{equation}
The coefficient $\delta^\varepsilon$ was obtained in \cite{KKLP24} as
\begin{equation}
\begin{aligned}
\delta^\eps
&=\left(1+\frac{3}{8}\frac{\texttt{h}}{\eps}\right)^{-2}
\left(\frac{\sts+(1+2\sqrt{3})\,\shs}{(8+3\sqrt{3})\,\shs}\right)
\frac{3 G_c}{16 W_{\texttt{ts}} \eps}
+\left(1+\frac{3}{8}\frac{\texttt{h}}{\eps}\right)^{-1}\frac{2}{5} ,
\end{aligned}
\label{delta-eps-final-h}
\end{equation}
where $\texttt{h}$ denotes the finite element size. { Note that the last term in the prescription for $c_{\texttt{e}}$ (\ref{cehat}) was introduced by Kumar et al. \cite{KRLP22} to improve the description of the compressive part of the strength surface from the governing PDEs (\ref{phase-field-equations}) for large values of localization lengths $\varepsilon$.}

The governing partial differential equations are solved using the finite element method. It is essential that the length scale $\varepsilon$ be fully resolved, so a fine mesh size is needed. We construct an unstructured mesh of size $\texttt{h}=\varepsilon/4$. Note that in some previous work \cite{molnar2024phase} studying the echelon crack formation, a very coarse mesh size of $\texttt{h}=\varepsilon$ was utilized, which is expected to affect the accuracy of the results. Also, in the previous work, the regularization length scale was tied to the material length scale. In contrast, $\varepsilon$ is a free parameter in our strength-constrained formulation and can be chosen to be as small as needed.  

The governing equations are solved iteratively with the fixed-point iteration method. The governing equation for the phase field must be solved subject to two constraints. The first constraint enforces that the phase field $v$ lies between 0 and 1. The second constraint enforces irreversibility of the phase field once a crack has formed. We make use of the penalty method to enforce both constraints. The details of the numerical implementation can be found in Kumar et al. \cite{KFLP18, KLDLP23}. We have also made available an open-source FEniCS implementation of the numerical scheme for this problem on GitHub \cite{KumarEchelonCode2026}. 

\begin{figure}[h]
	\centering
	\includegraphics[width=5.in]{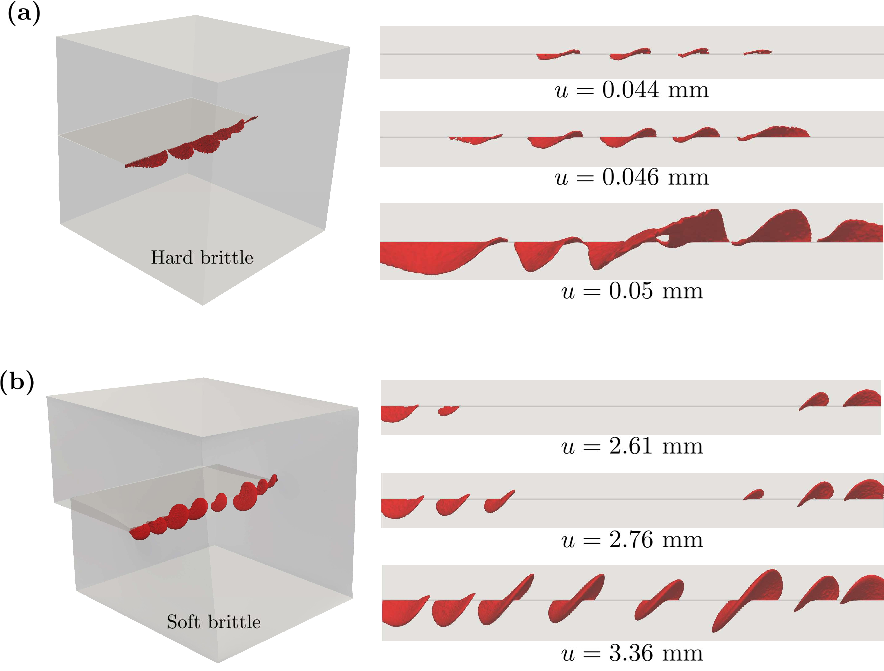}
	\caption{Simulations with the strength-constrained phase field model. Crack path contour for three values of applied displacement, $u$, for (a) graphite---a hard brittle material, and (b) PDMS---a soft brittle material.}\label{Fig3}
\end{figure}

\section*{Appendix C. Additional results on crack evolution and critical loads}

The results with the strength-constrained phase field model with graphite and PDMS for the smaller geometry (Fig.~\ref{Fig2}(a)) are shown in Fig.~\ref{Fig3}. Crack contours are shown at $v=0.1$ for three applied displacements in each case. See supplementary material, movies S1-S4 for the full evolution of cracks.

In graphite, four { child} cracks initially appear near the center, disconnected from one another but connected to the parent crack. They then reorient and propagate toward the lateral boundaries, with additional { child} cracks forming near the edges. As growth continues, the cracks eventually begin to bridge. In the nearly incompressible PDMS, { child} cracks first appear near the lateral surfaces and later in the center of the domain. They are inclined at a larger angle compared to those observed for graphite. This is because the ratio of compressive strength to tensile strength of PDMS is twice that of graphite. As shown before in Fig.~\ref{Fig4}(b), the orientation angle increases with the strength ratio.
Detailed analysis of crack spacing, reorientation, and coalescence lies outside the focus of the current work; studies addressing these aspects for hard brittle materials are available in the literature \cite{pham2017phase, santarossa2025configurational}.

\begin{figure}[h!]
	\centering
	\includegraphics[width=0.4\textwidth]{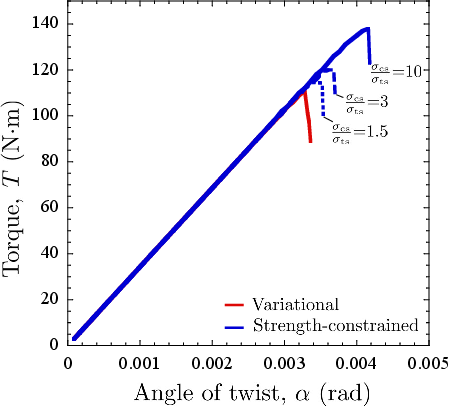}
	\caption{Comparison between the results for the resultant torque as a function of angle of twist for the classical variational and strength-constrained models in the torsion test.}\label{Fig7}
\end{figure}

Fig.~\ref{Fig4} evaluated the effect of the ratio of shear to tensile strength, $\sss / \sts$, on crack morphology.  
The strength ratio also influences the critical load at which cracks begin to form. Experimental validation of the strength-constrained model requires independent measurements of both tensile and shear strengths. However, prior experimental studies have reported only tensile strength, which makes direct validation difficult. The simplest way to measure shear strength is through an unnotched torsion test on a solid cylinder or a thin-walled circular tube \cite{kamarei2026nine}. In contrast, an unnotched tearing test is not suitable, as it involves a combination of tension and shear. Accordingly, in Fig.~\ref{Fig7},  we present results for the resultant torque as a function of the applied angle of twist for three values of $\scs / \sts$ for the test geometry shown in Fig.~\ref{Fig1}(d), which can support future experimental validation.

\bibliographystyle{elsarticle-num-names}
\bibliography{ref}

\end{document}

%% file: definitions.tex
\def\bfu{{\bf u}}

\def\bfy{{\bf y}}

\def\bfE{{\bf E}}

\def\bfI{{\bf I}}

\def\bfR{{\bf R}}
\def\bfS{{\bf S}}

\def\bfX{{\bf X}}
\def\bfF{{\bf F}}
\def\bfU{{\bf U}}

\def\eps{\varepsilon}

\def\e0{\varepsilon_0}

\def\s0{\sigma_0}

\def\sts{\sigma_{\texttt{ts}}}
\def\scs{\sigma_{\texttt{cs}}}
\def\sss{\sigma_{\texttt{ss}}}
\def\shs{\sigma_{\texttt{hs}}}

\DeclareMathAlphabet{\mathsfit}{T1}{\sfdefault}{\mddefault}{\sldefault}
\SetMathAlphabet{\mathsfit}{bold}{T1}{\sfdefault}{\bfdefault}{\sldefault}